# Kitaev interactions in the van der Waals antiferromagnet VBr$_3$

Zeyu Kao[1#], Yimeng Gu[1#], Yiqing Gu[1], Hao Zhang[1], Shiyi Zheng[1], Naoki Murai[2], Seiko Ohira-Kawamura[2], Jun Zhao[1,3,4*]

[1]State Key Laboratory of Surface Physics and Department of Physics, Fudan University, Shanghai 200433, China.

[2]Materials and Life Science Division, J-PARC Center, Tokai, Ibaraki 319-1195, Japan.

[3]Shanghai Research Center for Quantum Sciences, Shanghai 201315, China.

[4]Institute of Nanoelectronics and Quantum Computing, Fudan University, Shanghai 200433, China.

[#]These authors contributed equally to this work.
[*]e-mail: zhaoj@fudan.edu.cn

Abstract

Van der Waals materials hosting Kitaev interactions are promising platforms for exploring exotic quantum phenomena. Here, we report inelastic neutron scattering investigations of the van der Waals antiferromagnet VBr$_3$, which forms a honeycomb lattice structure at room temperature and exhibits zigzag-type magnetic order below 26.5 K. Our observations reveal distinctive low-energy spin excitations arising from Γ, Γ', and M' points, each featuring a spin gap of around 2.5 meV. The overall spin excitation spectra can be effectively described by a spin Hamiltonian incorporating substantial nearest-neighbor Kitaev and biquadratic interactions, along with Heisenberg interactions. Our findings not only establish VBr$_3$ as a new Kitaev magnet but also suggest that ligand engineering may provide a promising strategy to modulate Kitaev interactions, offering new opportunities for designing Kitaev materials with tailored quantum properties.

## 1. Introduction

The Kitaev interaction, an Ising-like interaction whose exchange easy axis depends on the spatial orientation of the exchange bond [1, 2], has attracted significant research interest in condensed matter physics. The spin-1/2 Kitaev model, whose ground state is the long-sought quantum spin liquid (QSL) [2], offers a novel pathway to QSL states through mechanisms beyond conventional geometric frustration [3]. The most widely studied Kitaev candidate materials are heavy transition metal (4d and 5d) Mott insulators on a honeycomb lattice, which exhibit a spin-orbit entangled $J_{eff} = 1/2$ state [4-6]. Well-known examples include the honeycomb iridates [7-9] and α-$RuCl_3$ [5, 10]. In addition to these, studies have also pointed to the potential for 3d transition metal compounds to exhibit Kitaev interactions [11-15]. However, rather than realizing a disordered QSL state, most of these candidate materials exhibit a zigzag-type antiferromagnetic order, which could be due to the presence of relatively strong non-Kitaev interactions.

Recently, inelastic neutron scattering experiments have revealed evidence of Kitaev interactions in the van der Waals honeycomb ferromagnet $VI_3$, which exhibits an effective spin of $S = 1$ at low temperatures [16]. This compound serves as a rare example of a high-spin Kitaev magnet, near the ferromagnetic instability. While the $S = 1$ Kitaev model is not exactly solvable, theoretical analyses and numerical calculations suggest that it may host a QSL state even in the presence of non-Kitaev interactions [17-19]. Calculations also suggest the presence of zigzag magnetic order near the QSL and ferromagnetic phases for $S = 1$ systems [18, 19], similar to the behavior observed in the $S = 1/2$ Kitaev materials.

Importantly, high-spin Kitaev systems may offer fundamentally distinct and richer physics compared to their $S = 1/2$ counterparts. The larger local Hilbert space introduces additional interaction channels, such as quadrupolar correlations and single-ion anisotropy. These degrees of freedom may enable novel magnetic orders and excitations, expanding the theoretical framework of Kitaev physics. In particular, it has been proposed that high-spin Kitaev magnets can host unconventional spin textures and topological phenomena not accessible in $S = 1/2$ systems [20], making them highly attractive for both fundamental research and potential applications.

However, the experimental exploration of high-spin Kitaev materials remains extremely limited. Even in the $S = 1/2$ systems with zigzag order, very scarce materials have been investigated using inelastic neutron scattering on single-crystalline samples to definitively determine the underlying magnetic interactions. Thus, elucidating magnetic interactions in vanadium-based materials with different magnetic ground states may offer a promising opportunity for achieving a more complete understanding of Kitaev materials.

Another intriguing aspect of vanadium trihalides is the potential influence of halogen atoms at the ligand sites on the magnetic interactions. Halogens connecting the vanadium atoms may influence spin-orbit coupling, which could, in turn, affect the nature and strength of the Kitaev interactions [21]. The influence of halogen atoms on the spin-orbit coupling and its interaction with the Kitaev mechanism remains largely unexplored experimentally, although recent theoretical studies have begun to offer valuable insights into this aspect [22, 23]. A deeper understanding of the relationship between the halogen-mediated spin-orbit coupling and the Kitaev interactions could open new avenues for the design of new Kitaev materials.

## 2. Materials and methods

In this context, we have recently synthesized large and high-quality $VBr_3$ single crystals using the chemical vapor transport method. Despite sharing a similar $BiI_3$-type crystal structure with $VI_3$ at room temperature, $VBr_3$ exhibits a markedly different magnetic ground state [24]. Our single-crystal neutron diffraction studies suggest that

VBr$_3$ exhibits zigzag-type antiferromagnetic order [25], which is similar to the extensively studied Kitaev magnet candidates such as α-RuCl$_3$ [26] and Na$_2$IrO$_3$ [27]. However, an alternative magnetic structure has also been proposed [28]. To elucidate the nature of the magnetic interactions and magnetic ground state, we conducted inelastic neutron scattering experiments on VBr$_3$ single crystals to investigate their magnetic excitation behavior.

Our inelastic neutron scattering experiments were conducted using the Cold-Neutron Disk-Chopper Spectrometer AMATERAS at Japan Proton Accelerator Research Complex (J-PARC) [29]. A total of 5.3 grams of crystals were coaligned for the experiments and the coaligned sample exhibited a sharp magnetic Bragg peak (0, 0.5, 1) (Fig. 1b), which is consistent with the presence of the zigzag-type magnetic order. No indications of magnetic order at (1, 0, 0.5) associated with the alternatively proposed magnetic structure [28] were detected in our coaligned crystals (Fig. S4 online). The inelastic neutron scattering data were collected at 3.5 K, 33 K and 120 K with incident neutron energies of 3.13, 7.74, 15.14, and 41.93 meV. Data visualization and analysis were performed using the Horace program, and the linear spin-wave simulations were carried out using the Su(n)ny package [30, 31].

3. Results

Constant-energy slices of spin excitations at various energies in VBr$_3$ are shown in the left panels of Figs. 2a–f. To capture the overall dispersion branches, the data are presented in the $L = 1$ plane, while the modulation of the scattering intensity along the out-of-plane ($L$) direction is shown in Fig. S2 (online). At low energies, magnetic excitations originating from the M' points are clearly visible, exhibiting an elliptical shape elongated along the longitudinal direction. This feature is consistent with the presence of the zigzag-type magnetic order. Additionally, strong magnetic excitations from the Γ point and weaker excitations from the Γ' point at similar energies are also observed. These excitations broaden with increasing energy and eventually merge, forming a kagome-shaped pattern above 10 meV (Fig. 2f).

The spin excitations along high-symmetry paths, illustrated in Figs. 3a–c, reveal several prominent features. Two branches of low-energy spin waves, observed between 2.5 and 5 meV (Figs. 3a and 3b), exhibit opposite dispersions: one branch disperses upward before bending downward, while the other disperses downward before bending upward. The peak energy of the first branch aligns with the lowest energy of the second branch at the high-symmetry points (0.5, 0, 1) and (1, 0, 1). At higher energies, around 9 meV, magnon band gap-like features emerge in multiple excitation branches (Figs. 3a–c). These distinct characteristics are further corroborated by constant-Q cuts (Figs. 4a–c). Additionally, the spin excitations display minimal dispersion along the *L* direction, indicating a quasi-two-dimensional nature of the magnetic correlations (Fig. S2 online). Below approximately 2.5 meV, spin gaps are observed at the Γ, Γ' and M' points (Figs. 3, 4a and 4b).

Attempts to model these features using the Heisenberg or XXZ models, or by incorporating Dzyaloshinskii-Moriya (DM) interactions, were unsuccessful (Figs. S5–S7 online), indicating that the spin dynamics in $VBr_3$ cannot be fully described by (quasi)-isotropic exchange interactions or antisymmetric exchanges. These discrepancies emphasize the need for a more comprehensive model that includes additional interactions, such as bond-dependent anisotropic exchange or higher-order terms, to more accurately explain the complex magnetic behavior observed in this material.

Interestingly, the magnetic excitations in $VBr_3$ are well-described by a model Hamiltonian that includes the Heisenberg interactions, a nearest-neighbor biquadratic interaction, a Kitaev interaction, and a single-ion anisotropy. The Hamiltonian for this model is presented in Eq. (1).

$$H = \sum_{\langle i,j\rangle_\gamma} \left[ J_1 \mathbf{S}_i \cdot \mathbf{S}_j + K_{\text{bq}}(\mathbf{S}_i \cdot \mathbf{S}_j)^2 + K S_i^\gamma S_j^\gamma \right] + \sum_{\langle\langle\langle i,l\rangle\rangle\rangle} J_3 \mathbf{S}_i \cdot \mathbf{S}_l$$

$$+ \sum_j A(\mathbf{S}_j \cdot \mathbf{n})^2$$

(1)

In this equation, $J_1$ and $J_3$ represent the nearest-neighbor and third-nearest-neighbor Heisenberg interactions, respectively, while $K_{\text{bq}}$ denotes the nearest-neighbor biquadratic interaction, and $K$ represents the Kitaev interaction. The variable $\gamma = x, y, z$ indicates the local axes for the bond-dependent Kitaev interaction. Different local axes for the bond-dependent Kitaev interaction are highlighted with different colors in Fig. 1a and Fig. S1b (online). $A$ represents the single-ion anisotropy term, and $\mathbf{n}$ indicates the easy-axis direction, which is consistent with the direction of the magnetic moment determined by neutron diffraction experiments. The direction of the magnetic moment lies nearly within the *ac*-plane, deviating from the *c*-axis by 36° (Fig. 1a) [25]. The simulation results are presented in the right panels of Fig. 2 and Fig. 3, providing a direct comparison with the corresponding experimental data. A finite broadening has been incorporated into the simulations to account for the combined effects of instrumental resolution, sample mosaic, and integration windows for the Horace scans. Our model accounts for three magnetic domains, each characterized by in-plane magnetic moments separated by 120° from one another [25]. Additionally, the $V^{3+}$ magnetic form factor was considered in the simulation. Our analysis reveals that the magnon band gap at 9 meV and the splitting of the lower magnon branches between 2.5 and 5 meV originate primarily from the Kitaev interaction, and its interplay with the biquadratic interaction and the third-nearest-neighbor Heisenberg interaction determines the size of the gap and splitting. The biquadratic term introduces effective anisotropy both along and across the zigzag chains, which modulates the spin-wave intensity distribution. The single-ion anisotropy predominantly governs the low-energy spin gap below around 2.5 meV. Off-diagonal exchange interactions were also considered during the fitting process; however, their inclusion did not substantially improve the fit quality or capture additional spectral features (Fig. S9 online). This may be related to the Br-V-Br angle being close to 90°, where the ideal Kitaev picture tends to yield weaker off-diagonal exchange [4, 32]. The influence of various exchange terms is illustrated in Fig. S8 (online). The parameters that provided the best fit to the data are as follows: $J_1 = -2.45$ meV, $J_3 = 0.43$ meV, $K_{bq} = -2.57$ meV, $K = -1.16$ meV, $A = -1.69$ meV.

Using these parameters, we computed the corresponding magnetic ground state by classical energy minimization. The optimized ordering wave vector and moment orientation agree with the neutron diffraction refinement (Fig. S10 online), confirming the self-consistency of the fitted Hamiltonian. A detailed comparison between the simulated and experimentally observed excitation dispersions reveals a remarkable agreement, which further validates our model (Figs. 2–4). Key features of the dispersion, including the opposite dispersions of the two low-energy branches and their merging at higher energies, are well reproduced in the simulations (Figs. 3d–f). The magnon band gap, which appears at energy transfer of about 9 meV with a magnitude of about 1.5 meV at high-symmetry points, is also captured (Figs. 3d–f, 4c). Moreover, our model successfully reproduces the full momentum dependence of the spin excitations across a broad range of momentum and energy (Fig. 2), including the six-point star-shaped pattern near the Γ point at low energies (Fig. 2a) and the kagome-like excitation pattern at higher energies (Fig. 2f). The anisotropic magnetic structure determined from diffraction experiments also supports the presence of ferromagnetic Kitaev interactions: the moments are oriented toward the midpoint of the octahedral edges rather than the vertices (Fig. S1b online), aligning with the local axes of the Kitaev interaction and indicating a ferromagnetic Kitaev interaction [25, 33]. Taken together, this comprehensive agreement between simulation and experiment underscores the critical role of Kitaev interactions in describing the complex magnetic behavior of $VBr_3$.

4. **Discussions and conclusion**

In contrast to $VI_3$, where two nearly degenerate orbital states have been observed and the high energy spin excitations may be influenced by the excited orbital state [16, 34, 35], $VBr_3$ exhibits a single set of spin-wave excitations, allowing for a more straightforward determination of the magnetic interactions, which are well described by our spin Hamiltonian model. Notably, the ordered magnetic moment in $VBr_3$ is approximately 0.89 $\mu_B$, significantly smaller than that suggested by the spin-only moment of the $V^{3+}$ ion ($S = 1$) [25]. This reduction suggests the presence of an unquenched orbital contribution, similar to those observed in $VI_3$ [16]. The prominent Kitaev interaction observed in $VBr_3$ is also consistent with the involvement of orbital degree of freedom in magnetic interactions in this system.

It is intriguing to compare $VBr_3$ with $VI_3$ and other Kitaev materials. Replacing iodine with bromine at the ligand sites in $VBr_3$ appears to modify the strength of the Kitaev interactions. The lighter bromine atoms reduce spin-orbit coupling compared to the heavier iodine atoms, which weakens the Kitaev interactions [22]. This substitution also alters other magnetic interactions, notably enhancing biquadratic interactions. The strength of higher-order magnetic interactions, such as the biquadratic exchange, is exquisitely sensitive to the local bonding geometry, particularly the metal-ligand-metal bond angle, which governs the orbital overlap between metal $d$ and ligand $p$ orbitals and thereby modulates the superexchange pathways. In edge-sharing octahedral systems, theory calculations have shown that when the metal-ligand-metal bond angle approaches 90°, the biquadratic exchange is abruptly enhanced [36]. In VBr$_3$, with a V-Br-V bond angle of about 90.5°, an enhanced biquadratic exchange is expected, which is consistent with our data. In VI$_3$, where the V-I-V bond angle is approximately 93.2°, this interaction should be smaller. First-principles calculations have indeed suggested the presence of such strong biquadratic interactions in VBr$_3$ with similar magnitude as our experiment [37].

As a result, the system's ground state shifts toward the zigzag antiferromagnetic configuration [38], suggesting a delicate balance between various magnetic interactions. While phase diagrams for Kitaev-Heisenberg systems have been extensively studied in theory [18, 19], the inclusion of biquadratic and single-ion anisotropy interactions, which are key features of high-spin systems, remains largely unexplored. These additional interactions may shift phase boundaries, stabilize new magnetic orders, or affect proximity to QSL phases. Future theoretical efforts that extend existing models to account for these effects would be particularly valuable, especially in the context of $S = 1$ systems, where quantum fluctuations are also pronounced.

From an experimental perspective, in Kitaev-Heisenberg materials such as α-RuCl$_3$, applying magnetic fields can suppress zigzag magnetic order and potentially induce a QSL state [39]. Additionally, the thermal Hall effect was also observed in α-RuCl$_3$ which could be linked to its proximity to the QSL state [39]. Given the significant Kitaev interactions in VBr$_3$ and its similar zigzag-type magnetic structure, it is particularly compelling to investigate whether VBr$_3$ exhibits hallmark responses of bond-directional magnets and potential signatures associated with proximity to a Kitaev spin-liquid regime, such as the field-direction-dependent suppression of zigzag order, anisotropy in the susceptibility, and transverse thermal transport. In particular, comprehensive experimental probes that go beyond zero-field spin-wave spectroscopy will be crucial, including field-dependent neutron scattering to track the suppression of magnetic order, spectral-weight redistribution and possible continuum formation, low-temperature specific heat and thermal conductivity to search for unconventional low-energy excitations, and local probes such as nuclear magnetic resonance/muon spin rotation/relaxation to examine the field evolution of static order and spin fluctuations.

Moreover, with considerable biquadratic interactions, together with the higher spin state, VBr$_3$ may host behavior beyond that of spin-1/2 counterparts [40, 41]. Indeed, recent theoretical calculations suggest the presence of novel topological phenomena, such as topological Hall effects, unique to high-spin Kitaev magnets [20]. Additionally, understanding how these effects relate to the behavior of ferromagnetic VI$_3$, where anomalous thermal Hall effects have also been observed [42], is of great interest. Therefore, VBr$_3$ could serve as an excellent material platform to study how the interplay between spin interactions and spin magnitude shapes emergent phenomena in Kitaev materials.

Finally, the contrasting magnetic ground states of VI$_3$ and VBr$_3$, despite their very similar crystal structures, highlight the potential of continuous halogen substitution from iodine to bromine in vanadium trihalides as an effective strategy for tuning Kitaev interactions without introducing significant disorder. By systematically varying the ligand chemistry, combined with the application of magnetic fields, gating, and strain, techniques particularly suitable for van der Waals materials, it may be possible to engineer Kitaev materials optimized for specific quantum phenomena, such as the long-sought QSL state. This tunability may enhance the versatility of Kitaev materials and offer opportunities for designing new quantum materials with tailored functionalities.

In summary, we have performed inelastic neutron scattering measurements on high-quality VBr$_3$ single crystals, successfully mapping out the magnetic excitations across a broad range of energy and momentum. The observed spin excitations are well described by a spin Hamiltonian that incorporates significant nearest-neighbor Kitaev interactions, biquadratic interactions, and Heisenberg exchange interactions. By demonstrating that Kitaev interactions can be both tuned and observed, our work establishes vanadium-based materials as a new class of Kitaev magnets. This provides a valuable platform to comprehensively explore the field of high-spin Kitaev magnets, with important implications for the design and engineering of quantum materials.

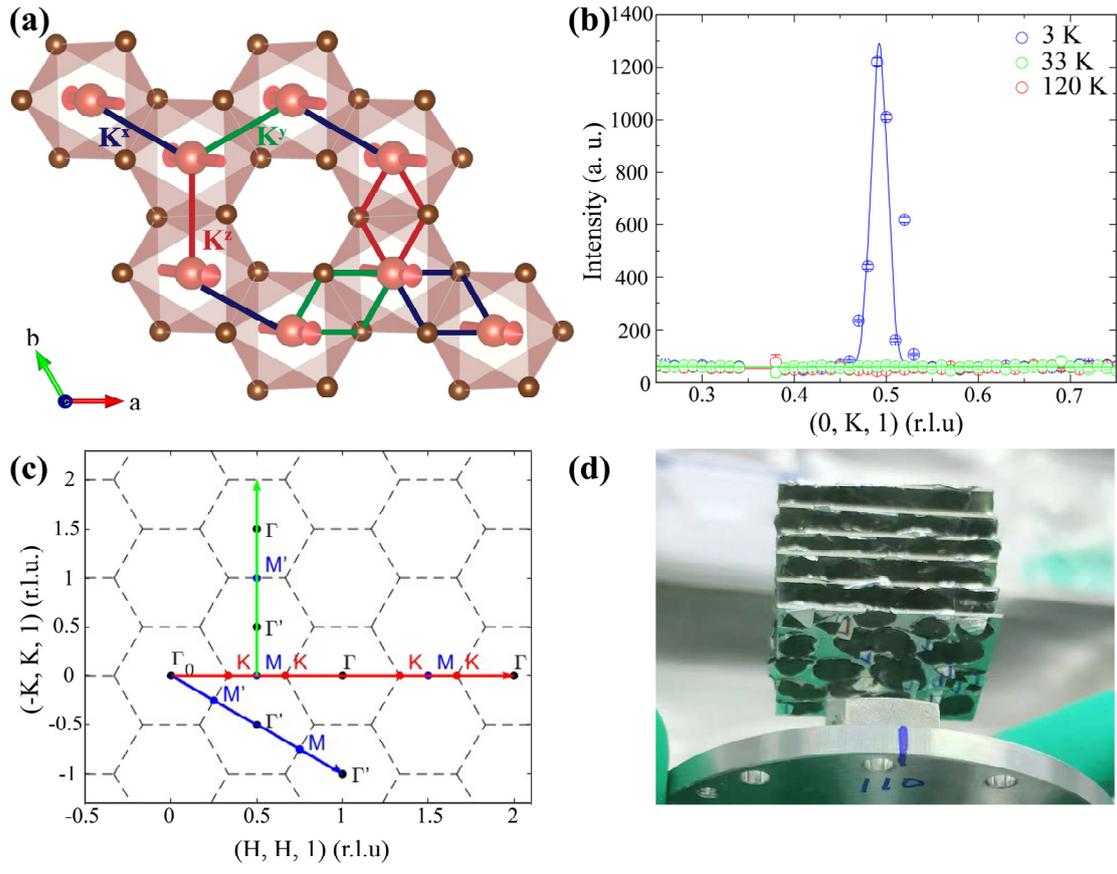

Fig. 1. (a) Crystal structure of VBr$_3$, where V$^{3+}$ ions form a honeycomb lattice. Each V$^{3+}$ ion is coordinated by six Br$^-$ ions, creating an edge-sharing octahedral geometry that underpins the bond-dependent Kitaev interactions. The $K_x$ (blue), $K_y$ (green), and $K_z$ (red) represent the three types of bonds with orthogonal Ising axes in the Kitaev model. Red arrows depict the zigzag antiferromagnetic ground state, with the zigzag chains aligned along the $a$-axis. (b) Magnetic Bragg peak from neutron scattering measurements on the coaligned sample. The sharp magnetic Bragg peak (0, 0.5, 1) reflects the good quality and reasonably small mosaic spread of the samples, and it disappears above the Néel temperature. The elastic signal was integrated over an energy window of ±0.1 meV around the elastic line. (c) Reciprocal lattice of VBr$_3$, showing three high-symmetry directions. (d) A photograph of coaligned VBr$_3$ crystals used for inelastic neutron scattering experiments.

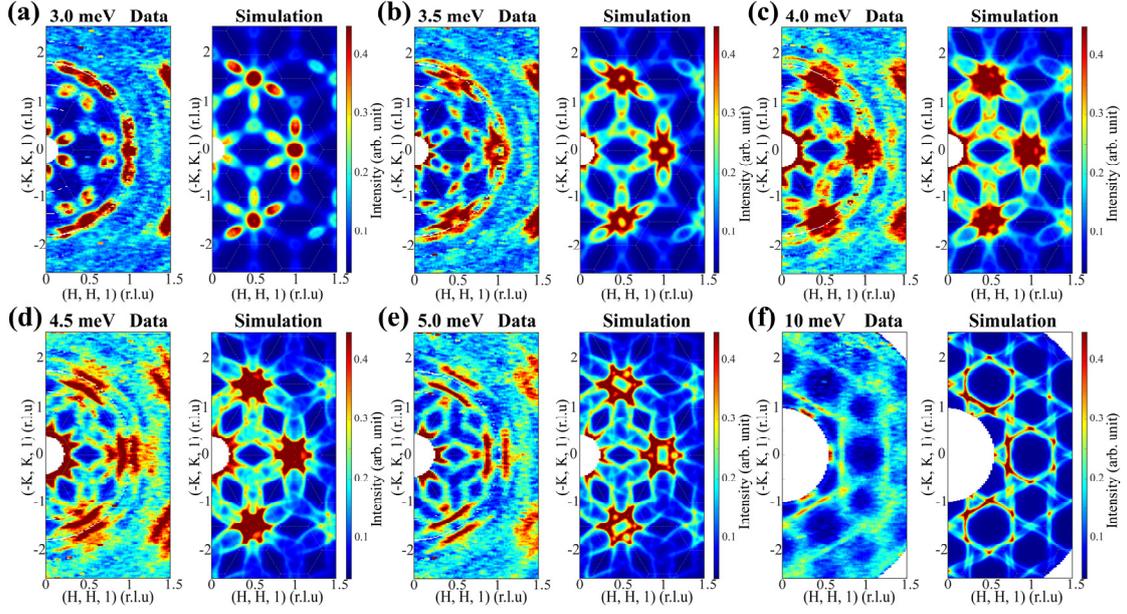

Fig. 2. In-plane constant energy slices of spin excitations in VBr$_3$. (a)–(f) The left panels show experimental data with energy transfers of $E$ = 3.0, 3.5, 4.0, 4.5, 5.0, and 10 meV, integrated over $E \pm 0.15$ meV and $0.7 \leq L \leq 1.3$. The corresponding simulation results are displayed in the right panels.

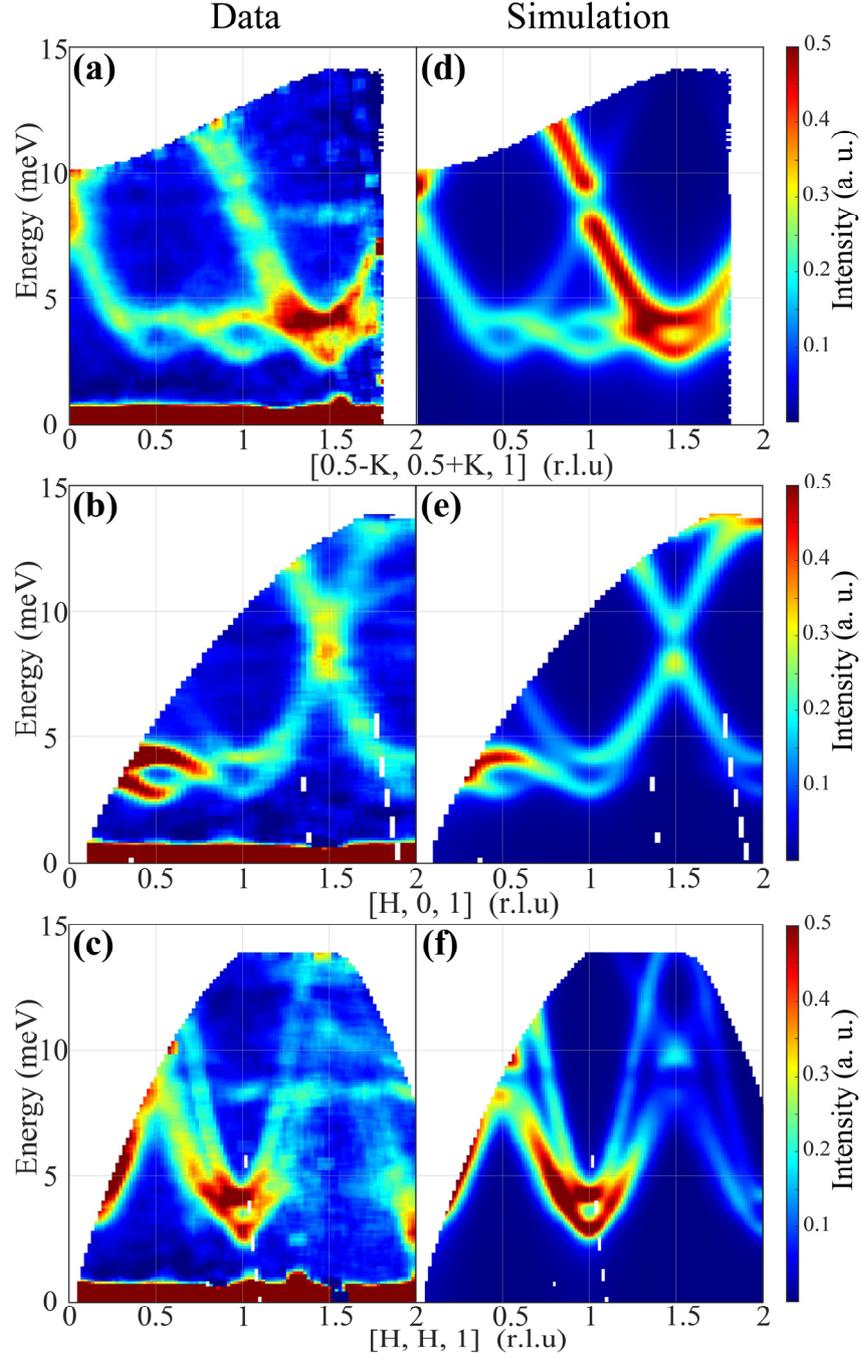

Fig. 3. Spin wave dispersions in VBr3 along high-symmetry directions. (a)–(c) Measured magnetic excitations along the green, blue, and red paths in Fig. 1c. Data were collected at T = 3 K with an incident neutron energy of $E_i$ = 15 meV. The integration range is ±0.1 along the in-plane vertical direction and ±0.1 along the $L$ direction near $L$ = 1. (d)–(f) Simulated magnetic excitations based on the Heisenberg-biquadratic-Kitaev model, corresponding to the experimental data shown in (a)–(c). The experimental excitation spectra exhibit slightly broader features than the simulations, which may result from the finite mosaic spread of the coaligned crystals or intrinsic damping of the magnon excitations. Additionally, we observe flat

excitations around 8 meV in panels (a) and (e), as well as a weak dispersive mode above 10 meV at $H > 1.5$ in panel (c), which are not reproduced in our simulations. These features are attributed to phonon modes, as evidenced by their increasing intensities with higher momentum transfer (Q) and their persistence at temperatures above the Néel temperature (Fig. S3 online).

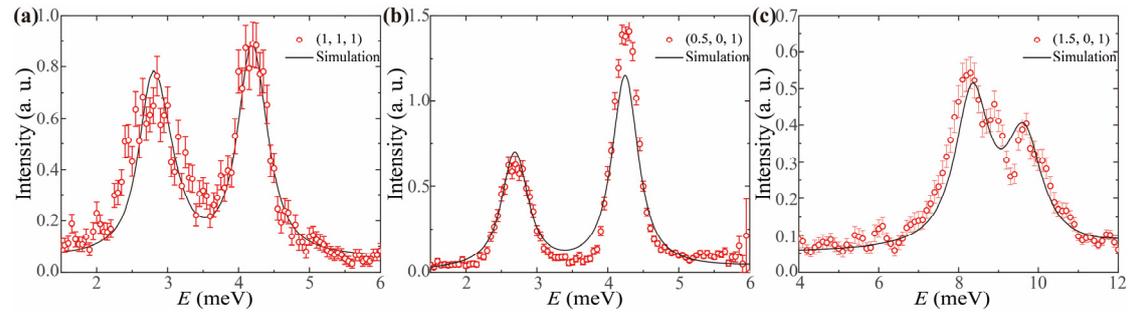

Fig. 4. Constant Q cuts at (a) the Γ point, (b) the M' point, and (c) the M point. The data are integrated over ±0.03 reciprocal-lattice units in the in-plane direction and over ±3 along the $L$ direction to improve statistics. Solid lines correspond to cuts generated from the best-fit $S(Q,\omega)$.

# Supplementary Materials for
# Kitaev interactions in the van der Waals antiferromagnet VBr$_3$

**Details of neutron scattering experiments**

For the inelastic neutron scattering experiment, we selected 194 high-quality single crystals of VBr$_3$, with a total mass of 5.3 grams. These crystals were co-aligned within the HK0 plane, achieving a mosaic spread of approximately 5°. The experiment was conducted using the AMATERAS time-of-flight neutron spectrometer at the Japan Proton Accelerator Research Complex (J-PARC) [1].

Considering the energy scale of VBr$_3$, along with the neutron flux and energy resolution of the spectrometer, we selected incident energies of 3.13, 7.74, 15.14, and 41.93 meV, yielding respective energy resolutions of 0.06, 0.23, 0.60, and 2.5 meV at the elastic line. Given the structural phase transition of VBr$_3$ at $T_S$ = 90.4 K and the antiferromagnetic phase transition at $T_N$ = 26.5 K, measurements were performed at 3 K, 33 K, and 120 K.

Although the in-plane honeycomb lattice of VBr$_3$ undergoes slight distortion upon cooling through the structural phase transition at $T_S$ = 90.4 K, the resulting lattice distortions are minimal. Consequently, we approximate the lattice as a honeycomb structure at low temperature and adopt a hexagonal unit cell (*R-3*, *a* = *b* = 6.37 Å, *c* = 18.38 Å, *α* = *β* = 90°, *γ* = 120°) for data processing in the HORACE program [2].

For clearer visualization, the intensity maps generated with HORACE were subjected to a light nearest-neighbor smoothing, a standard practice for time-of-flight neutron data. We applied 'hat' windowing function to perform weighted averaging on the data. We used the weakest level of smoothing, where only a 3×3 matrix surrounding each data point was involved in the averaging. The weighting matrix used was [1, 2, 1; 2, 4, 2; 1, 2, 1]. This smoothing approach effectively reduces single-pixel noise while preserving the intrinsic features of the data and producing smoother edges.

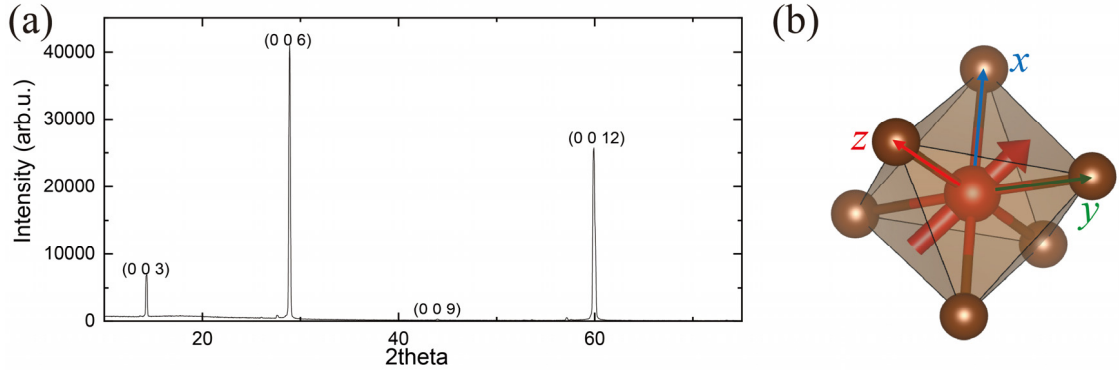

Fig. S1. (a) X-ray diffraction pattern along the *ab* plane on a representative VBr$_3$ single crystal used for the inelastic neutron scattering experiment. (b) Illustration of VBr$_6$ octahedron and the magnetic moment direction (bold red arrow). The arrows with marks represent the local axes of bond-dependent Kitaev interaction.

**Dispersion of magnetic excitations along *L***

During the cooling process, a structural phase transition occurs, which breaks the threefold rotational symmetry. This process is completely random, leading to the formation of three distinct domains. Given that the inelastic experiments utilized a large amount of sample, it can be statistically assumed that the volume fractions of the three crystal domains are equal. Therefore, in our dispersion fitting, we have taken into account three crystal domains that are mutually separated by 120°.

The nearly dispersionless magnetic excitations along the *L* direction confirm the strong two-dimensional nature of the spin correlations in VBr$_3$ (Fig. S2). The intensity of the lower branch at (0.5, 0) (Fig. S2c) exhibits modulation with varying *L*, which is accurately reproduced in our simulation (Fig. S2d). This branch originates from one of the 120° twins, corresponding to an actual wave vector $Q = (0, -0.5, L)$. It arises from the vibration of the magnetic moment perpendicular to the *ac* plane, corresponding to the $S_{yy}$ component of the spin-spin correlation function. At $L = 0$, the vibration is parallel to $Q$, resulting in zero intensity. As $L$ increases, the vertical component of the vibration intensifies, leading to an increase in intensity. Therefore, we present $L = 1$ rather than $L = 0$ in the main text figures to capture all dispersion branches, ensuring comprehensive coverage of the *ab* plane and maintaining high data quality.

**Temperature dependence of excitations**

Fig. S3 displays the excitation spectra measured at various temperatures. As the temperature increases above the Néel temperature, the intensity of magnetic excitations gradually diminishes. In contrast, the phonon excitations observed in Fig. S3a and S3b, such as those around 8 meV between $H = 1$ and $H = 2$, remain robust and persist even at high temperatures. Similarly, the excitations observed in Fig. S3c and S3d at around 10~11meV from $H = 1.5$ to $H = 2$ also originate from phonon. This temperature-dependent behavior confirms the magnetic nature of the low-energy excitations. The increased background observed at higher temperatures is likely due to enhanced phonon scattering from several sources, including the aluminum sample holder, the Cytop glue used for coalignment, helium exchange gas, and the sample itself.

**Magnetic Bragg peaks associated with the zigzag order in coaligned crystals**

Figure S4 presents a constant energy slice at $E = 0$ meV. Prominent magnetic Bragg peaks corresponding to the zigzag antiferromagnetic order are clearly observed at the wave vector ***k*** = (0, 0.5, 1) [3]. No indications of magnetic peaks at ***k*** = (1, 0, 0.5) associated with the proposed alternative magnetic structure [4] were detected in our coaligned crystals.

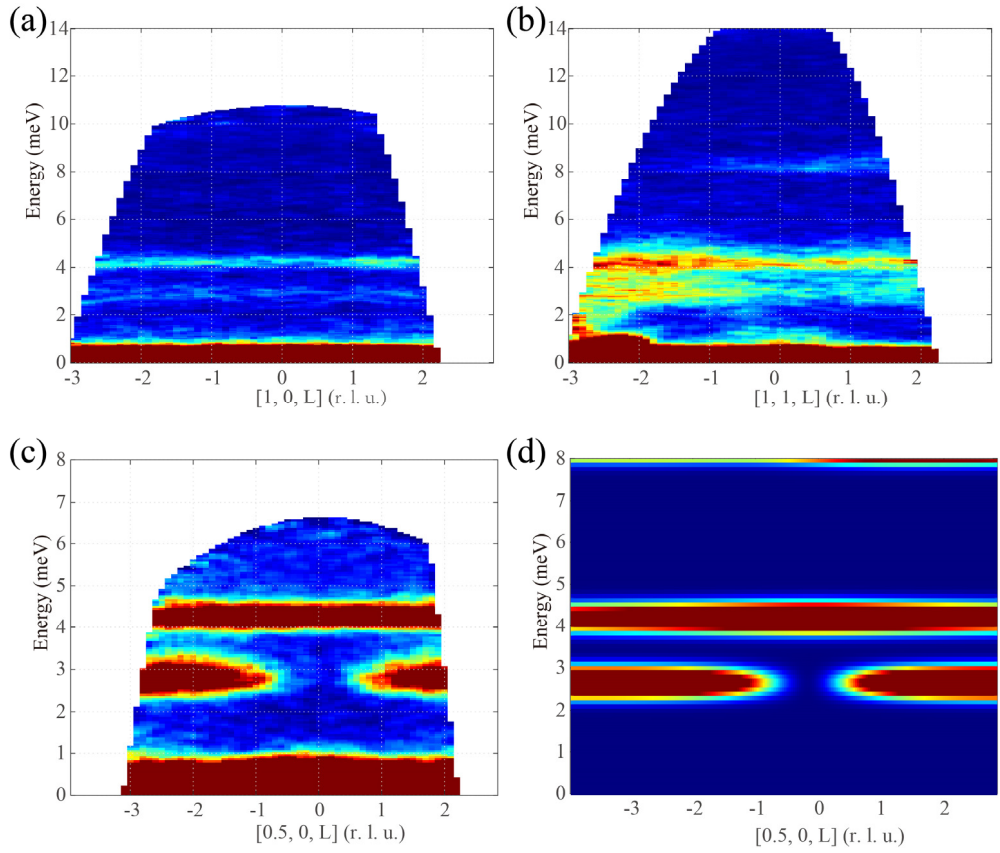

Fig. S2. Dispersions of magnetic excitations along the *L* direction in VBr$_3$ measured at 3 K. (a) Dispersion along the (1, 0, *L*) direction. (b) Dispersion along the (1, 1, *L*) direction. (c) Dispersion along the (0.5, 0, *L*) direction. (d) Simulated magnetic excitations corresponding to (c), based on the Heisenberg-biquadratic-Kitaev model described in the main text.

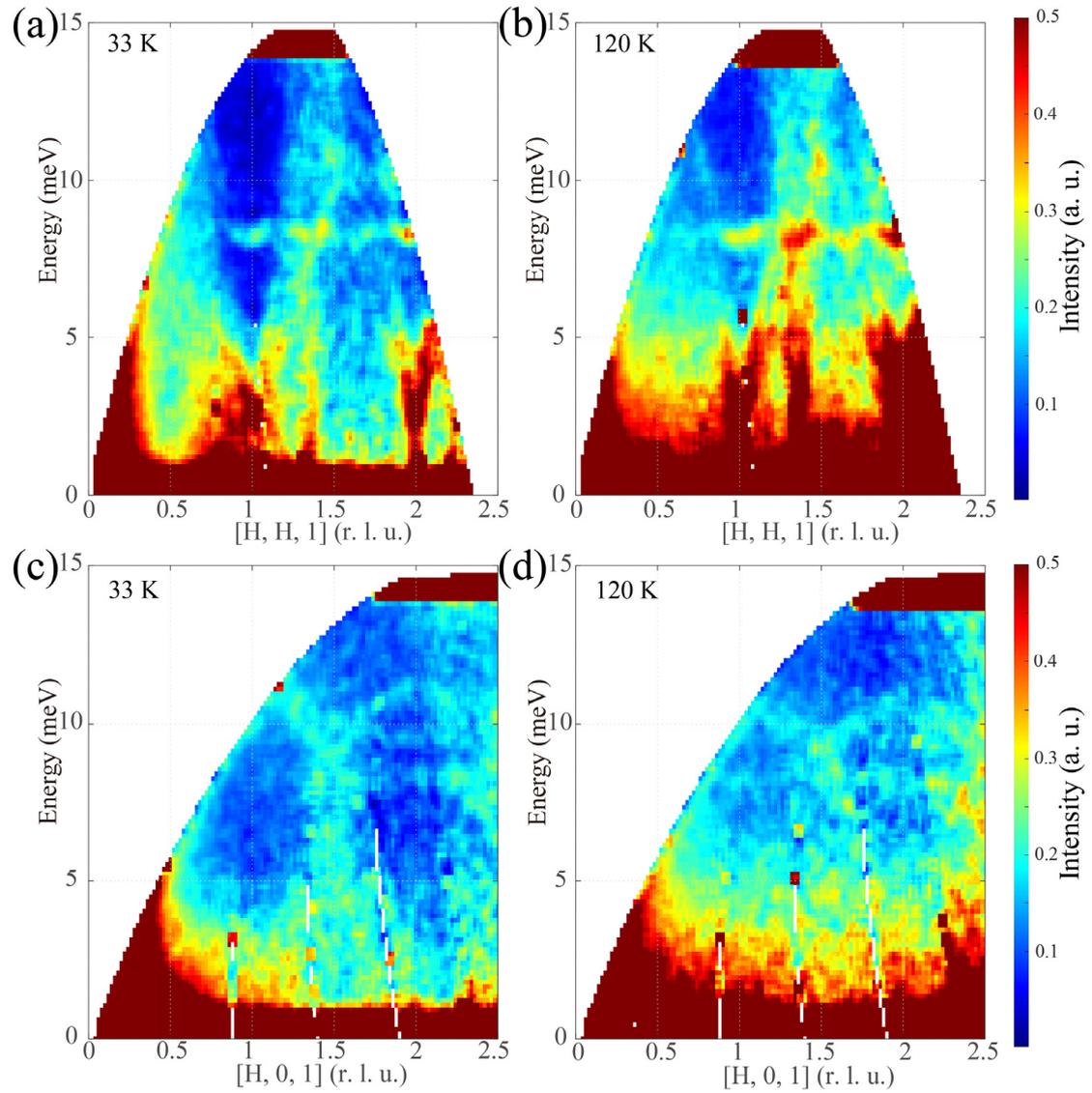

Fig. S3. Excitation spectra along high symmetry directions at various temperatures in VBr$_3$. (a) Excitation spectra along the (*H*, *H*) direction at 33 K. (b) Excitation spectra along the (*H*, *H*) direction at 120 K. (a) Excitation spectra along the (*H*, 0) direction at 33 K. (d) Excitation spectra along the (*H*, 0) direction at 120 K.

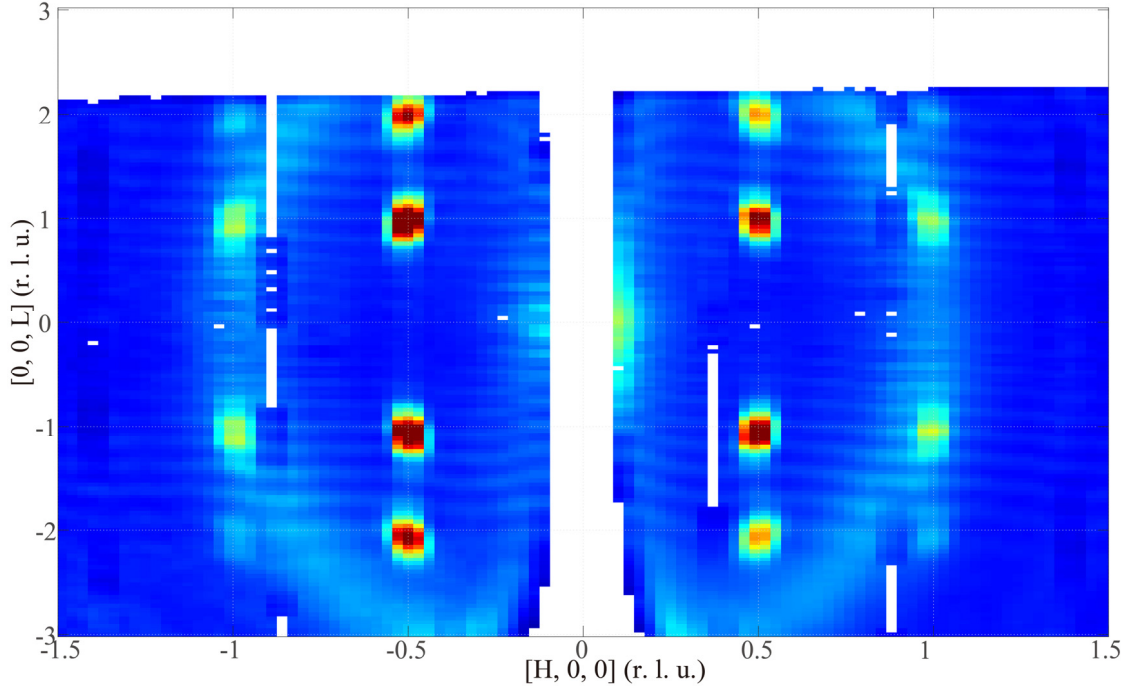

Fig. S4. Constant-energy slice at $E = 0$ meV and at 3 K reveals prominent magnetic Bragg peaks corresponding to the zigzag antiferromagnetic order at the wave vector $\boldsymbol{k} = (0, 0.5, 1)$ in our coaligned crystals.

**Comparison of different models**

The XXZ honeycomb model takes the form,

$$H = \sum_{\langle i,j \rangle} \left[ J_1 \left( S_{xi}S_{xj} + S_{yi}S_{yj} + \lambda S_{zi}S_{zj} \right) + K_{bq}\left( \boldsymbol{S}_i \cdot \boldsymbol{S}_j \right)^2 \right] + \sum_{\langle\langle\langle i,l \rangle\rangle\rangle} J_3 \boldsymbol{S}_i \cdot \boldsymbol{S}_l$$

$$+ \sum_j A\left( \boldsymbol{S}_j \cdot \boldsymbol{n} \right)^2$$

Here, $\lambda$ contributes to the XXZ-type exchange anisotropy in the nearest-neighbor interaction. We keep the biquadratic term and the third-nearest-neighbor Heisenberg interaction in the model. $A$ represents the single-ion anisotropy, while $\boldsymbol{n}$ denotes the direction of the easy axis.

The symmetry-allowed DM interaction model takes the form,

$$H = \sum_{\langle i,j \rangle} \left[ J_1 \mathbf{S}_i \cdot \mathbf{S}_j + K_{bq}(\mathbf{S}_i \cdot \mathbf{S}_j)^2 \right] + \sum_{\langle\langle i,k \rangle\rangle} \mathbf{D}_{ik} \cdot (\mathbf{S}_i \times \mathbf{S}_k) + \sum_{\langle\langle\langle i,l \rangle\rangle\rangle} J_3 \mathbf{S}_i \cdot \mathbf{S}_l$$

$$+ \sum_j A(\mathbf{S}_j \cdot \mathbf{n})^2$$

The term $\mathbf{D}_{ik} \cdot (\mathbf{S}_i \times \mathbf{S}_k)$ represents the symmetry-allowed next-nearest-neighbor DM interaction, where the $\mathbf{D}_{ik}$ vector is perpendicular to the next-nearest-neighbor-bond, which is constrained by symmetry analyses. Given that, $\mathbf{D}_{ik}$ can be decomposed into two components: an in-plane component $D_\parallel$ and an out-of-plane component $D_\perp$.

We use the Su(n)ny package to generate the spin excitation spectrum, with results presented in Figs. S5–S7. When utilizing the XXZ model, no significant splitting of the lower branches (highlighted by the red rectangular box in Fig. S5a) was observed, in stark contrast to the predictions of the Kitaev model. Furthermore, the XXZ model fails to reproduce the magnon band gap around 9 meV within the red box of Fig. S5b, a feature that is accurately captured by the Kitaev model. As for the symmetry-allowed DM interaction model, the simulation also fails to reproduce the overall spin wave dispersion observed in our experimental data.

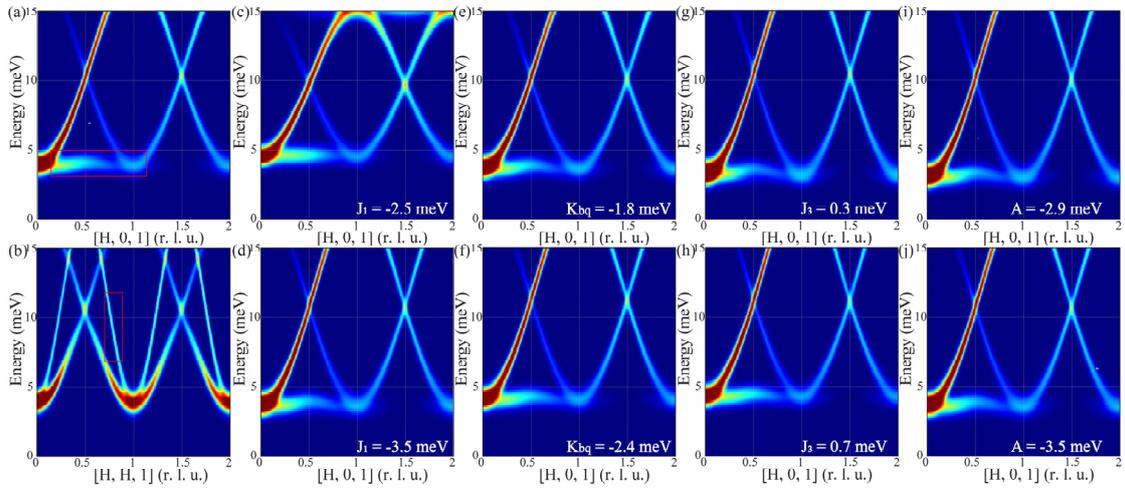

Fig. S5. Simulated spin waves for the XXZ model along (a) the [H, 0, 1] direction and

(b) [H, H, 1] direction. Simulation parameters: $J_1 = -3.2$ meV, $\lambda = 1.4$, $K_{bq} = -2$ meV, $J_3 = 0.5$ meV, $A = -3.2$ meV. (e–j) Simulations using different parameters. Except for the parameters specified in the figure, all other parameters are identical to those used in (a–b).

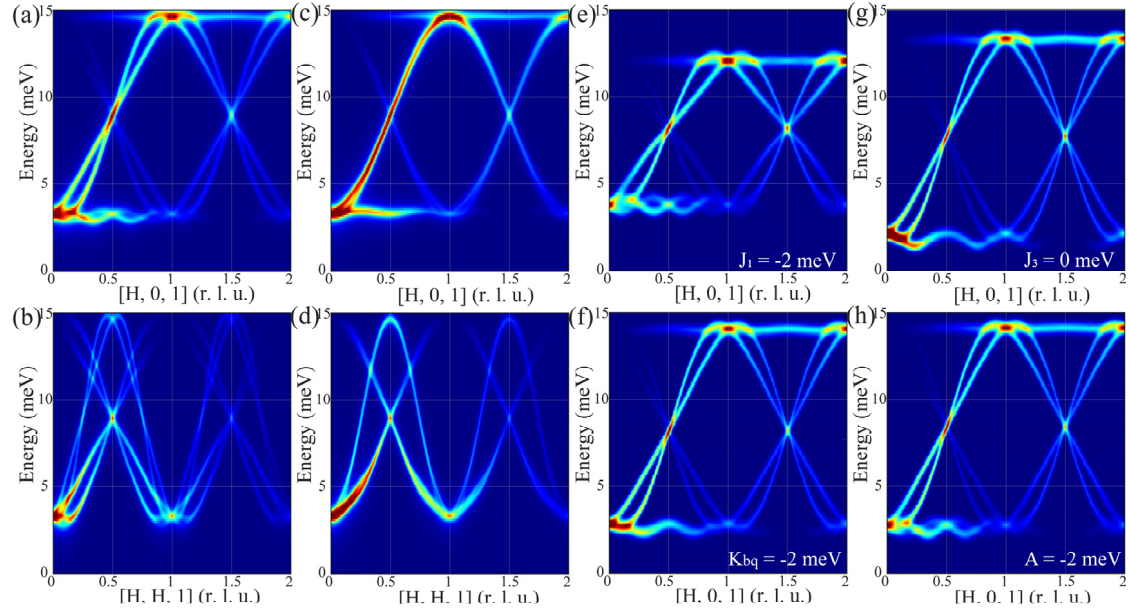

Fig. S6. Simulated spin waves with different DM interaction parameters. (a) Spin wave dispersions along [H, 0, 1] with DM components $D_\parallel = 0$, $D_\perp = 0.5$ meV. (b) Spin wave dispersions along [H, H, 1] with DM components $D_\parallel = 0$, $D_\perp = 0.5$ meV. (c) Spin wave dispersions along [H, 0, 1] with $D_\parallel = 1$ meV, $D_\perp = 0$. (d) Spin wave dispersions along [H, H, 1] with $D_\parallel = 1$ meV, $D_\perp = 0$. Simulation parameters: $J_1 = -2.8$ meV, $K_{bq} = -2.57$ meV, $J_3 = 0.43$ meV, $A = -2.5$ meV. (e–h) Simulations using different parameters. Except for the parameters specified in the figure, all other parameters are identical to those used in (a).

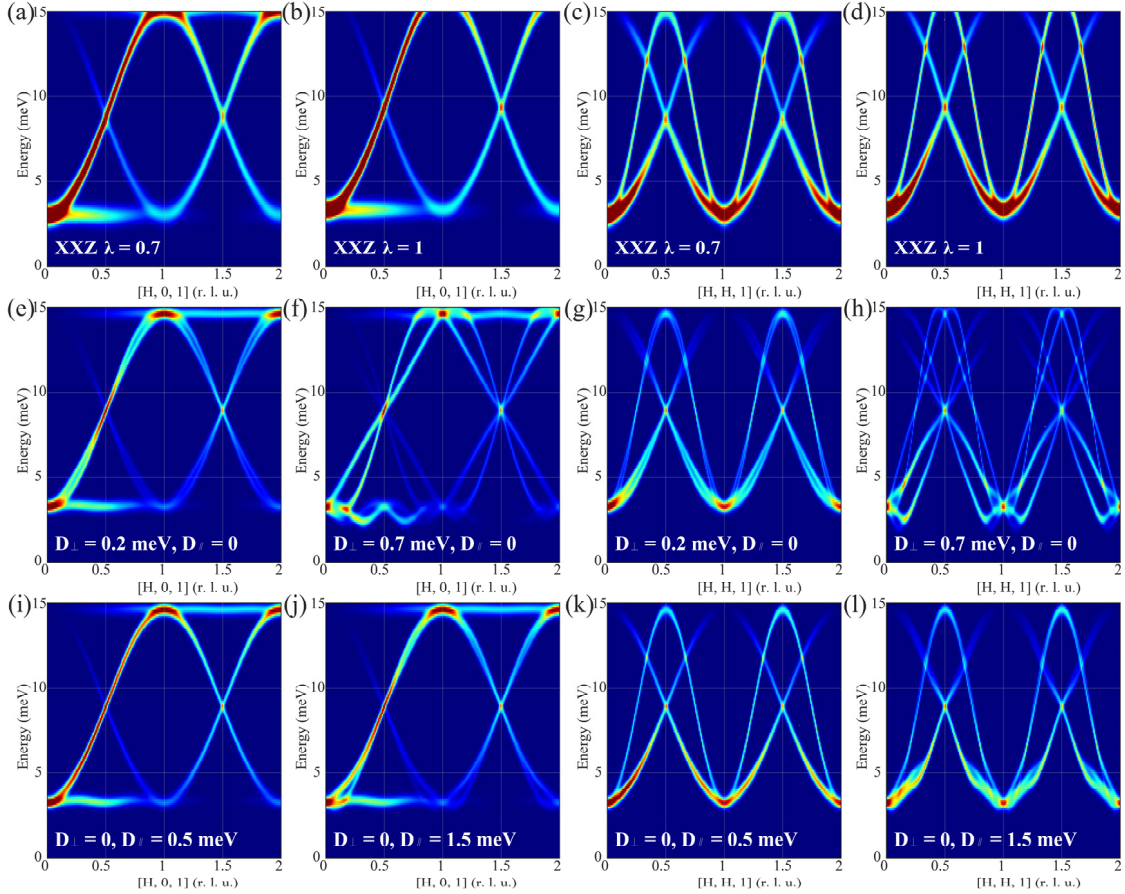

Fig. S7. Simulation results of different models with different parameters. Except for the parameters specified in the figure, all other parameters are identical to those used in Fig. S5–6.

**The influence of different parameters on the fitting results**

Initially, we conducted a comprehensive sweep of the parameter space. Among all the sweep results, we identified the simulated outcomes that closely matched our experimental data. Finally, through fine-tuning of these matched results, we derived the final parameters described in the manuscript. This systematic process ensures that, within the broad parameter space, our fitting results exhibit a certain degree of uniqueness and correctness.

To clarify the influence of individual parameters in our model, we present the simulation results in Fig. S8. The $J_1$ term establishes the overall framework of the spin wave spectrum (Fig. S8a–b). $J_3$, $K$, $K_{bq}$ terms primarily influence the shape of the lower two branches around [0.5, 0] between 2.5 and 5 meV (Fig. S8c–h). The gap opening at 9 meV and the associated branch splitting arise from the Kitaev term (Fig. S8g–h). The biquadratic term modulates the spin-wave intensity distribution (Fig. S8e–f). The single-ion anisotropy primarily tunes the low-energy spin gap, with relatively minor effects on other features (Fig. S8i–j).

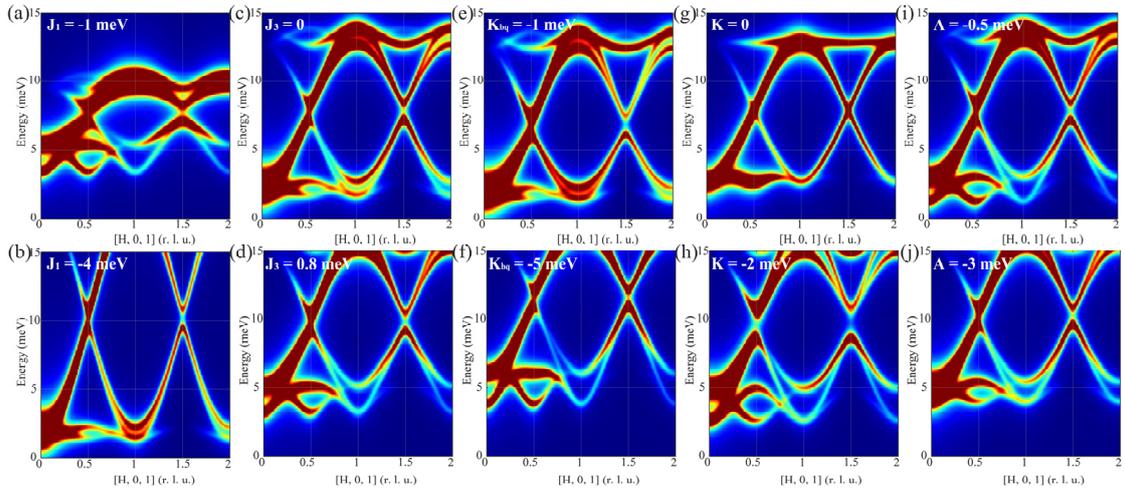

Fig. S8. Simulation results using different parameters. Each figure differs from the main text in one parameter. (a) $J_1 = -1$ meV. (b) $J_1 = -4$ meV. (c) $J_3 = 0$. (d) $J_3 = 0.8$ meV. (e) $K_{bq} = -1$ meV. (f) $K_{bq} = -5$ meV. (g) $K = 0$. (h) $K = -2$ meV. (i) $A = -0.5$ meV. (j) $A = -3$ meV. Best-fit parameters: $J_1 = -2.45$ meV, $J_3 = 0.43$ meV, $K_{bq} = -2.57$ meV, $K = -1.16$ meV, $A = -1.69$ meV.

Considering the off-diagonal terms, the Hamiltonian can be written in the following form:

$$H = \sum_{\langle i,j \rangle \in \alpha\beta(\gamma)} \left[ J_1 \mathbf{S}_i \cdot \mathbf{S}_j + K_{bq}(\mathbf{S}_i \cdot \mathbf{S}_j)^2 + K S_i^\gamma S_j^\gamma + \Gamma\left(S_i^\alpha S_j^\beta + S_i^\beta S_j^\alpha\right) \right.$$

$$\left. + \Gamma'\left(S_i^\alpha S_j^\gamma + S_i^\gamma S_j^\alpha + S_i^\beta S_j^\gamma + S_i^\gamma S_j^\beta\right) \right] + \sum_{\langle\langle\langle i,l \rangle\rangle\rangle} J_3 \mathbf{S}_i \cdot \mathbf{S}_l$$

$$+ \sum_j A(\mathbf{S}_j \cdot \mathbf{n})^2$$

Where $\langle i,j \rangle \in \alpha\beta(\gamma)$ represents the nearest-neighbor bonds and the Ising axes of Kitaev interaction ($\alpha$, $\beta$, $\gamma$ represents three mutually perpendicular axes on different nearest-neighbor bonds). As shown in Fig. S9, introducing off-diagonal terms does not lead to a better description of the inelastic neutron scattering data.

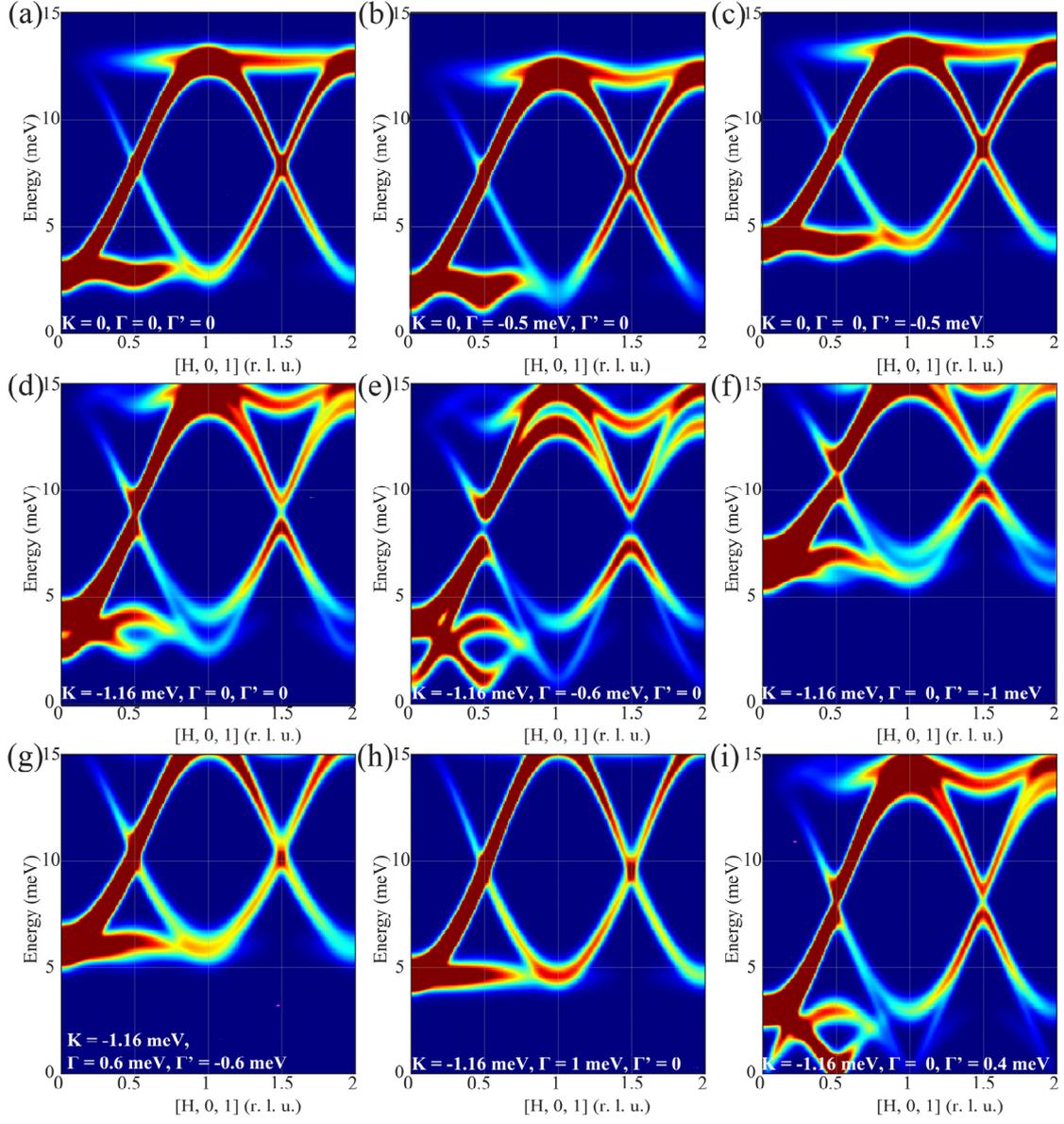

Fig. S9. Simulation results with off-diagonal terms. Except for the parameters specified in the figure, other parameters are: $J_1 = -2.45$ meV, $J_3 = 0.43$ meV, $K_{bq} = -2.57$ meV, $A = -1.69$ meV.

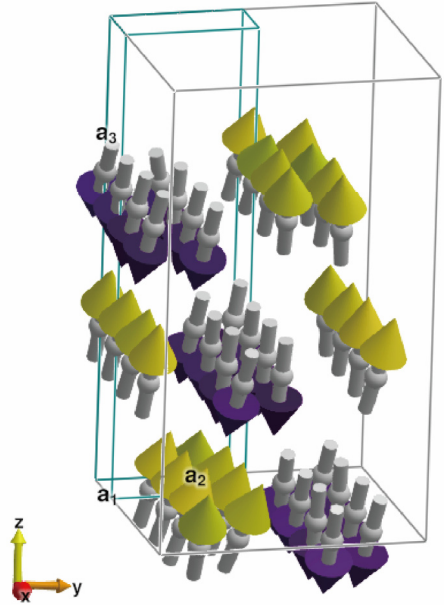

Fig. S10. Classical ground state from energy minimization of the fitted spin Hamiltonian (parameters as in the main text). The calculated ordering wave vector and moment orientation agree with those obtained from neutron diffraction [3].